\documentclass[twocolumn,apl,epsf,superscriptaddress]{revtex4}
\usepackage{amsmath}
\usepackage{amsfonts}
\usepackage{epsfig}
\usepackage{epsf}
\usepackage{array}
\usepackage{color}
\usepackage{ulem}

\begin{document}

\title{Charge transfer in iridate-manganite superlattices\footnote{
Copyright  notice: This  manuscript  has  been  authored  by  UT-Battelle, LLC under Contract No. DE-AC05-00OR22725 with the U.S.  Department  of  Energy.   
The  United  States  Government  retains  and  the  publisher,  by  accepting  the  article  for  publication, 
acknowledges  that  the  United  States  Government  retains  a  non-exclusive, paid-up, irrevocable, world-wide license to publish or reproduce the published form of this manuscript, 
or allow others to do so, for United States Government purposes.  
The Department of Energy will provide public access to these results of federally sponsored  research  in  accordance  with  the  DOE  Public  Access  Plan 
(http://energy.gov/downloads/doe-public-access-plan)}}

\author{Satoshi Okamoto}
\altaffiliation{okapon@ornl.gov}
\affiliation{Materials Science and Technology Division, Oak Ridge National Laboratory, Oak Ridge, Tennessee 37831, USA}
\author{John Nichols}
\affiliation{Materials Science and Technology Division, Oak Ridge National Laboratory, Oak Ridge, Tennessee 37831, USA}
\author{Changhee Sohn}
\affiliation{Materials Science and Technology Division, Oak Ridge National Laboratory, Oak Ridge, Tennessee 37831, USA}
\author{Soy Yeun Kim}
\affiliation{Center for Correlated Electron Systems, Institute for Basic Science (IBS), Seoul 08826, Republic of Korea}
\affiliation{Department of Physics \& Astronomy, Seoul National University, Seoul 08826, Republic of Korea}
\author{Tae Won Noh}
\affiliation{Center for Correlated Electron Systems, Institute for Basic Science (IBS), Seoul 08826, Republic of Korea}
\affiliation{Department of Physics \& Astronomy, Seoul National University, Seoul 08826, Republic of Korea}
\author{Ho Nyung Lee}
\affiliation{Materials Science and Technology Division, Oak Ridge National Laboratory, Oak Ridge, Tennessee 37831, USA}

\begin{abstract}
Charge transfer in superlattices consisting of SrIrO$_3$ and SrMnO$_3$ is investigated using density functional theory. 
Despite the nearly identical work function and non-polar interfaces between  SrIrO$_3$ and SrMnO$_3$, rather large charge transfer was experimentally reported at the interface between them. 
Here, we report a microscopic model that captures the mechanism behind this phenomenon, 
providing a qualitative understanding of the experimental observation.  
This leads to unique strain dependence of such charge transfer in iridate-manganite superlattices. 
The predicted behavior is consistently verified by experiment with soft x-ray and optical spectroscopy. 
Our work thus demonstrates a new route to control electronic states in non-polar oxide heterostructures. 
\end{abstract}

\maketitle

\date{\today }

Electron density is one of the most important parameters controlling electronic phases in strongly correlated electron systems. 
As a milestone in condensed matter physics, 
high critical temperature superconductivity was discovered in Cu-based oxides by doping carriers into Mott insulating states \cite{Bednorz1986}. 
This triggered an improvement in crystal synthesis techniques, leading to the discovery of a number of novel spin, charge and orbital states in complex oxide materials \cite{Imada1998}. 
Thin film growth techniques have also improved dramatically \cite{Izumi2001,Ohtomo2002}. 
In Ref. \cite{Ohtomo2002}, Ohtomo {\it et al.} demonstrated atomically sharp interfaces between two insulating titanates with a metallic behavior.
Such metallic interfaces led to the concept of electronic reconstruction originally proposed for K-doped C$_{60}$ systems \cite{Hesper2000,Okamoto2004a}. 
One of the important aspects of the electronic reconstruction is that 
the valence state of constituent ions in such heterostructures can significantly differ from the corresponding valence state in bulk systems as a result of the electron transfer within the heterostructures. 
Such electron transfer can be manipulated by the polar discontinuity \cite{Ohtomo2004} 
or by the difference in the work functions \cite{Yunoki2007}. 
The polar discontinuity was previously discussed in the context of III-V semiconductor heterostructures  \cite{Harrison1978}.  
In this case, the discontinuity often leads to the atomic reconstruction because 
it is significantly more challenging to change the valence state than for transition-metal elements.

Thus, hetero-structuring is expected to become a fascinating route to explore novel electronic states in complex systems 
by controlling the valence state or the carrier density without introducing disorder intrinsic to chemically doped bulk crystals. 
There have been a large number of reports along this direction \cite{Ahn2006}, and we anticipate this field will continue to grow rapidly \cite{Hwang2012}. 
Because of the renewed interest in the relativistic spin-orbit coupling (SOC) with correlations, 
iridium-based systems have started to attract significant interest \cite{Kim2008}. 
There have already appeared a number of theoretical predictions for novel phenomena \cite{Wang2011,Xiao2011,Okamoto2013,Okamoto2014} 
and experimental studies \cite{Hirai2015,Matsuno2015,Matsuno2016} using perovskite-type iridates.

Recently, superlattices of SrIrO$_3$ (SIO) and SrMnO$_3$ (SMO) were epitaxially grown and their transport and magnetic properties were reported  \cite{Nichols2016}. 
Among a number of characteristics, one of the most striking results was a quite large charge transfer
of $\sim 0.5$ electrons per unit cell from SIO to SMO regions. 
Because of the nearly identical work function between SIO and SMO, 5.05~eV and 4.99~eV, respectively \cite{Shimizu2005}, 
one would naively expect that the amount of transferred electrons to be rather small and even in the opposite direction, i.e., from SMO to SIO. 

In this paper, we investigate the charge transfer between SIO and SMO in their superlattices using density functional theory (DFT). 
Our DFT results show a reasonable agreement with the experimental report in Ref. \cite{Nichols2016}. 
We also construct a phenomenological model to understand the microscopic mechanism of the surprising charge transfer at SIO/SMO interfaces. 
This phenomenological model is based on molecular orbitals formed at the interface and naturally predicts unique strain dependence of the charge transfer, 
which we confirm by DFT calculations and verify experimentally. 

The charge transfer across non-polar interfaces was also discussed between GdTiO$_3$ and perovskite nickelates \cite{Grisolia2016}. 
The amount of charge redistribution is controlled by the covalent character of the transition metal/oxygen bonds. 
The molecular orbital formation provides an alternative means to control the charge transfer across non-polar interfaces 
by utilizing the interfacial interaction between constituent systems.

The DFT calculations were performed with the generalized gradient approximation and projector augmented wave 
approach \cite{Blochl1994} 
as implemented in the Vienna {\it ab initio} simulation package (VASP) \cite{Kresse1999,Kresse1996}. 
For Ir, Mn and O, standard potentials were used (Ir, Mn and O in the VASP distribution) 
and for Sr, a potential, in which semicore $s$ and $p$ states are treated as valence states, is used (Sr$_{sv}$). 
To account for strong correlation effects \cite{Dudarev1998}, we included the local $U$ for Ir and Mn $d$ states; 
$U=2$~eV for Ir $d$ \cite{Arita2012} and $U=3$~eV for Mn $d$ \cite{Picozzi2007}.  
We performed two sets of calculations. 
First, the structural optimization was performed without spin polarizations and the SOC  
using the doubled unit cell with the experimental in-plane lattice constant of SrTiO$_3$, $a=b=3.905 \times \sqrt{2}$~{\AA}, 
a $4 \times 4 \times 4$ $k$-point grid, and an energy cutoff of 500 eV. 
%
Optimized crystal structures were achieved when forces on all the atoms were $<0.01$~eV/{\AA}. 
Subsequently, we determined magnetic order with finite SOC. 
This procedure finds lattice parameters of bulk SMO ($a=3.814$~{\AA}) and SIO ($a=3.984$~{\AA}) close to the experimental ones  
$a=3.80$~{\AA} \cite{Takeda1974} and $a=3.94$~{\AA} \cite{Longo1971,Zhao2008,Biswas2014}, respectively,  
validating our theoretical approaches. 
%
Second, we further relaxed the fractional atomic coordinates with the spin polarization and the SOC. 
During this process, we fixed the in-plane and out-of-plane lattice constants as obtained without spin polarizations and the SOC.  
If the full structural optimization including out-of-plane lattice constant is performed with the spin polarizations instead, 
the effect of compressive strain would be overestimated 
-- the lattice constant of bulk SMO has been determined to be more than 1~\% larger than the experimental value.\cite{Lee2010} 
In the following, we will present the two sets of theoretical results, the first data followed by the second data shown in parenthesis. 
The two sets of results are qualitatively consistent and, thus, the physical trends are robust.

%

\begin{figure}[tbp]
\begin{center}
\includegraphics[width=0.9\columnwidth, clip]{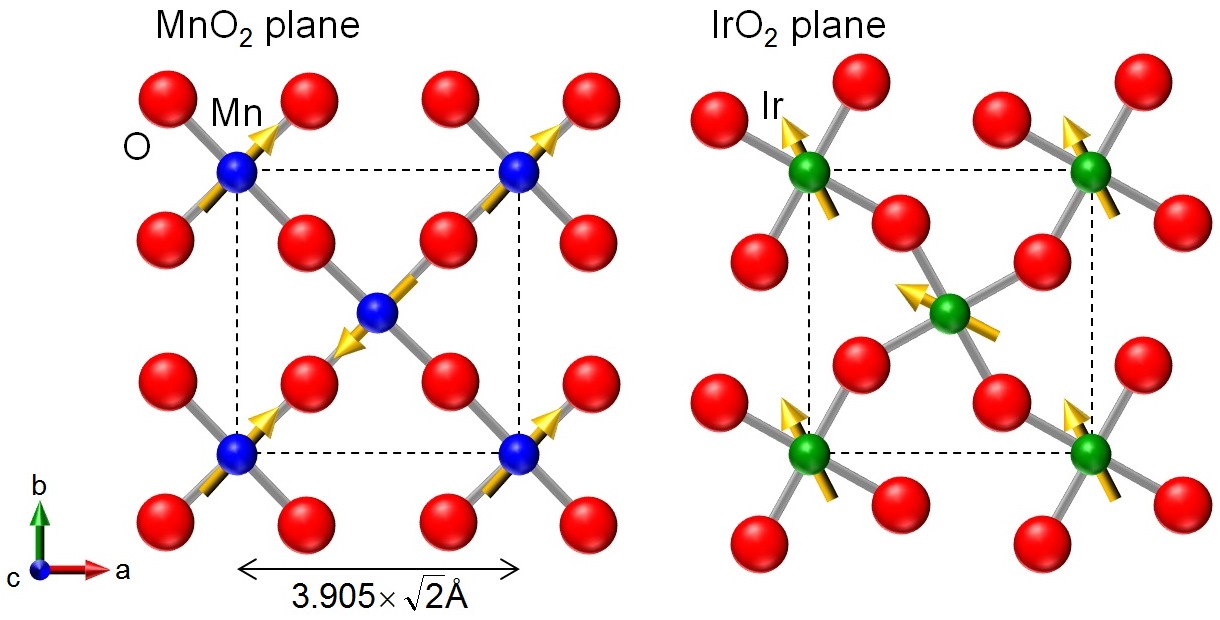}
\caption{Structure of a [SrIrO$_3$]$_1$[SrMnO$_3$]$_1$ superlattice optimized without magnetism and SOC. 
Yellow arrows indicate the ordered spin moments obtained with additional SOC. 
Their size does not reflect the actual amplitude.}
\label{fig:M1I1}
\end{center}
\end{figure}

Figure \ref{fig:M1I1} shows the structure of a [SIO]$_1$[SMO]$_1$ superlattice optimized without magnetism and the SOC, 
and magnetic ordering is subsequently analyzed under this structure. 
One can notice stronger structural distortion in SIO than SMO, i.e., larger rotation of the IrO$_6$ octahedron than the MnO$_6$. 
The rotation angle of the IrO$_6$ octahedron is found to be $\alpha \sim 15.2^\circ (14.4^\circ)$, which is substantially larger than 
that in bulk Sr$_2$IrO$_4$ $\alpha \sim 11^\circ$ \cite{Crawford1994}. 
The larger distortion arises from the mismatch ($\sim 0.9$~\%) between SIO ($a = 3.94$~{\AA} \cite{Longo1971,Zhao2008,Biswas2014}) 
and STO ($a = 3.905$~{\AA}), the substrate material used in Ref. \cite{Nichols2016}. 

The size of the ordered moment is found to be 0.12 (0.13)~$\mu_B$ at Ir sites and 3.13 (3.02)~$\mu_B$ at Mn sites. 
The former is nearly identical to the DFT estimate on Sr$_2$IrO$_4$ with a smaller $U=1.6$~eV \cite{Liu2015}. 
However, our canting angle is found to be extremely large $72.0^\circ (58.5^\circ)$ compared with that in Sr$_2$IrO$_4$ ($14.4^\circ$ \cite{Liu2015}),  
and the nearest-neighboring coupling is almost ferromagnetic.
Note that the canting angle $90^\circ$ corresponds to the case where the ordered moments are parallel to the diagonal direction. 
This does not only come from the larger rotation angle of the IrO$_6$ octahedra than in Sr$_2$IrO$_4$, 
but also comes from the reduced Ir $d$ occupancy, by which the antiferromagnetic interactions between neighboring Ir sites are suppressed. 
On the other hand. Mn moments remain to be coupled antiferromagnetically pointing along the nearest-neighboring Mn-O directions. 
Therefore, the Ir moments have large contributions to a net magnetic moment along the diagonal direction (see Fig. \ref{fig:M1I1}). 
We found that the Mn moment 3.13 (3.02)~$\mu_B$  is enhanced from the bulk value that we determine to be 2.77~$\mu_B$. 
This enhancement is attributed to the electron transfer from SIO to SMO. 
Assuming the transfered electrons enter into Mn majority spin bands, the enhanced magnetization corresponds to charge transfer of 0.36 (0.25) electrons per Mn.

In principle, one could estimate the transfered electrons in the VASP output file. 
However, the values obtained in this procedure have large ambiguity because of the overlap with ligand O $p$ states. 
When the electron density is increased on a Mn site, the Mn-O bond length is normally increased. 
This increase results in the reduced Mn-O overlap and the underestimation of the charge density. 
In the current study, we measure the difference in Mn moment from the bulk SMO value. 
The latter (2.77~$\mu_B$) already involves minority spins due to the hybridization with O $p$ states.   
However, as will be discussed in detail later, 
since the transfered electrons predominantly enter into the Mn $3z^2-r^2$ orbital with their spins parallel to $t_{2g}$ spins due to the strong  Hund coupling, 
the change in the Mn moment is a good measure for the transfered electrons.

We have also analyzed [SIO]$_3$[SMO]$_1$, [SIO]$_1$[SMO]$_3$  and [SIO]$_3$[SMO]$_3$ superlattices, 
and all the results are summarized in Table \ref{tab:summary}. 
In  [SIO]$_3$[SMO]$_1$ superlattice, the ordered Mn moment  is enhanced to 3.18 (3.56)~$\mu_B$, corresponding to 0.41 (0.79) electron per Mn. 
%
In  [SIO]$_{1}$[SMO]$_3$ and [SIO]$_{3}$[SMO]$_3$ superlattices, 
0.15 (0.14) and 0.30 (0.27) electrons, respectively, are found to be transfered to a SMO layer, which is adjacent to a SIO layer. 
Thus, we notice that the electron transfer from a SIO region to a SMO region is enhanced by increasing the SIO thickness $m$ when the SMO thickness $n$ is fixed. 
This indicates that the charge screening length in SIO is longer than 1 unit cell, 
while the precise determination remains difficult because SIO undergoes magnetic to non-magnetic transition with increasing its thickness.\cite{Matsuno2015} 
The difference between  [SIO]$_m$[SMO]$_1$ superlattices and  [SIO]$_m$[SMO]$_3$ superlattices 
is ascribed to the different geometry as discussed later.

\begin{table}[tbp]
\caption{DFT results on the magnitude of electron transfer from a SIO region to a SMO layer, which is adjacent to a SIO layer, in [SIO]$_m$[SMO]$_n$ superlattices. 
Inside brackets are the results of additional structural relaxation with the spin polarization and the SOC. 
Note that this does not affect our conclusion at all. }
\label{tab:summary}
\newcommand{\lw}[1]{\smash{\lower2.0ex\hbox{#1}}}
\begin{center}
\begin{tabular}{cccc}
$m$ & $n$ & magnetic moment [$\mu_B$] & electron transfer $\Delta N$ \\
\hline
1 & 1 & 3.13 (3.02) & 0.36 (0.25) \\
3 & 1 & 3.18 (3.56) & 0.41 (0.79) \\
1 & 3 & 2.92  (2.91) & 0.15 (0.14) \\
3 & 3 & 3.07 (3.04) & 0.30 (0.27) 
\end{tabular}
\end{center}
\end{table}

Our DFT calculations and the experimental measurement in Ref. \cite{Nichols2016} have shown rather unique results, 
especially strong electron transfer from SIO to SMO in spite of the nearly identical work function between the two materials. 
First, we considered the possibility of oxygen non-stoichiometry.  
However, since the sum the the Ir and Mn valence states is approximately constant, this scenario seems to be quite unlikely. 
Thus, in order to understand this unique behavior,
we propose the use of a molecular orbital picture \cite{Okamoto2010}. 
Although this is a phenomenological approach, it can provide a physically transparent image. 
Since SMO has the $t_{2g}^3$ electron configuration with $S=3/2$, the doubly degenerate $e_g$ states are empty. 
On the other hand, SIO has the $t_{2g}^5$ configuration. 
Because of strong SOC, the $t_{2g}^5$ manifold is split into fourfold degenerate $J_{eff}=3/2$ states, which are fully filled, 
and the twofold degenerate $J_{eff}=1/2$ states, which are half filled, 
leaving empty $e_g$ states that are far above the Fermi level.

\begin{figure}[tbp]
\begin{center}
\includegraphics[width=0.7\columnwidth, clip]{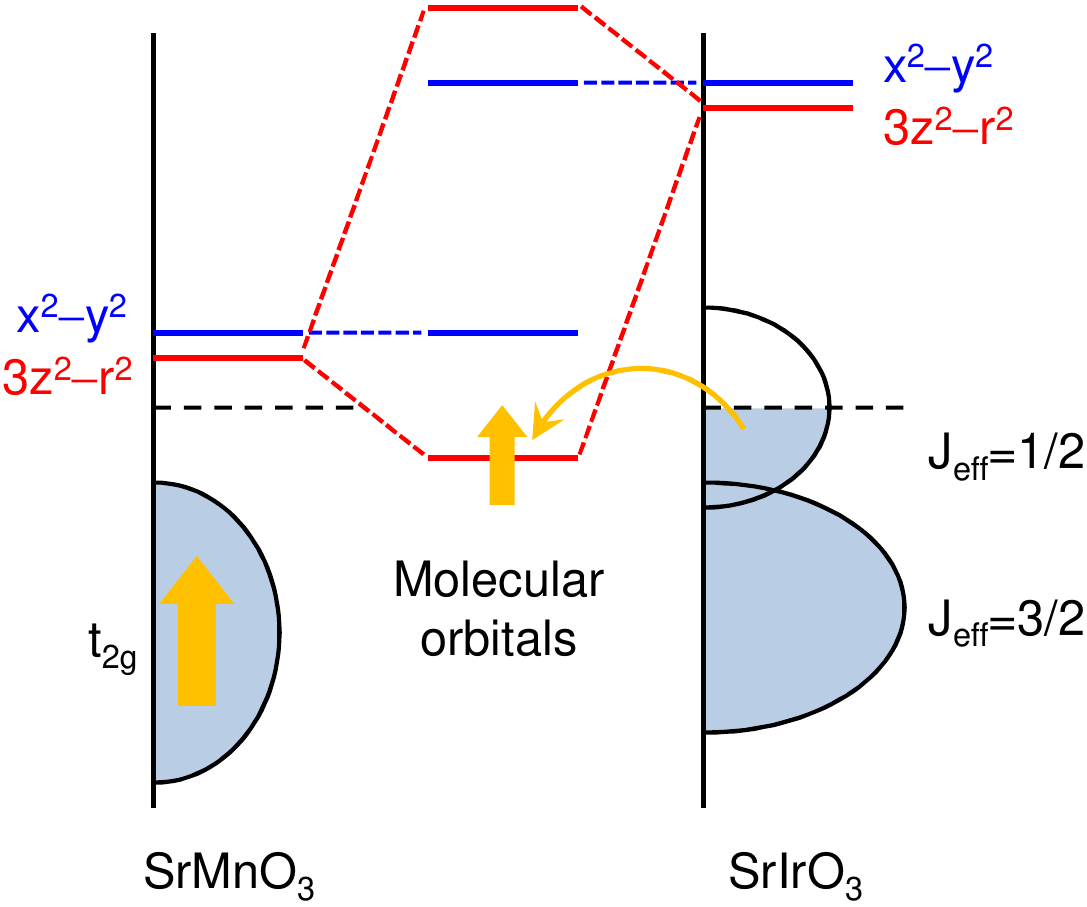}
\caption{Schematic density of states of SrMnO$_3$ and SrIrO$_3$, and molecular orbitals formed at an interface between them.
The $t_{2g}$ states are shown with finite band widths, while $e_g$ states are shown as localized levels for simplicity. 
For SrMnO$_3$, only majority spin states are shown as indicated by the arrow. 
For SrIrO$_3$, spins are not explicitly considered.}
\label{fig:molecular}
\end{center}
\end{figure}

Considering these electron configurations and the similar work functions between them, 
the single particle density of states (DOS) of SIO and SMO would align as shown in Fig. \ref{fig:molecular}. 
Because of the strong overlap of $3z^2-r^2$ orbitals between neighboring Mn and Ir sites, they are expected to form molecular orbitals. 
Bonding orbitals have a strong Mn character and are located below the lower Mn $3z^2-r^2$ level, 
and antibonding orbitals have a stronger Ir character and are located above the higher Ir $3z^2-r^2$ level. 
On the other hand, the formation of bonding and antibonding molecular orbitals is much weaker for $x^2-y^2$ orbitals because of the weaker hybridization along the $z$ direction. 
As a result, the bonding $3z^2-r^2$ orbitals become lower than the original Fermi level. 
Because of the smaller or zero gap in SIO, electrons are transfered  from the $J_{eff}=1/2$ manifold of SIO into the bonding $3z^2-r^2$ orbitals, i.e., from SIO to SMO. 

\begin{figure}[tbp]
\begin{center}
\includegraphics[width=0.8\columnwidth, clip]{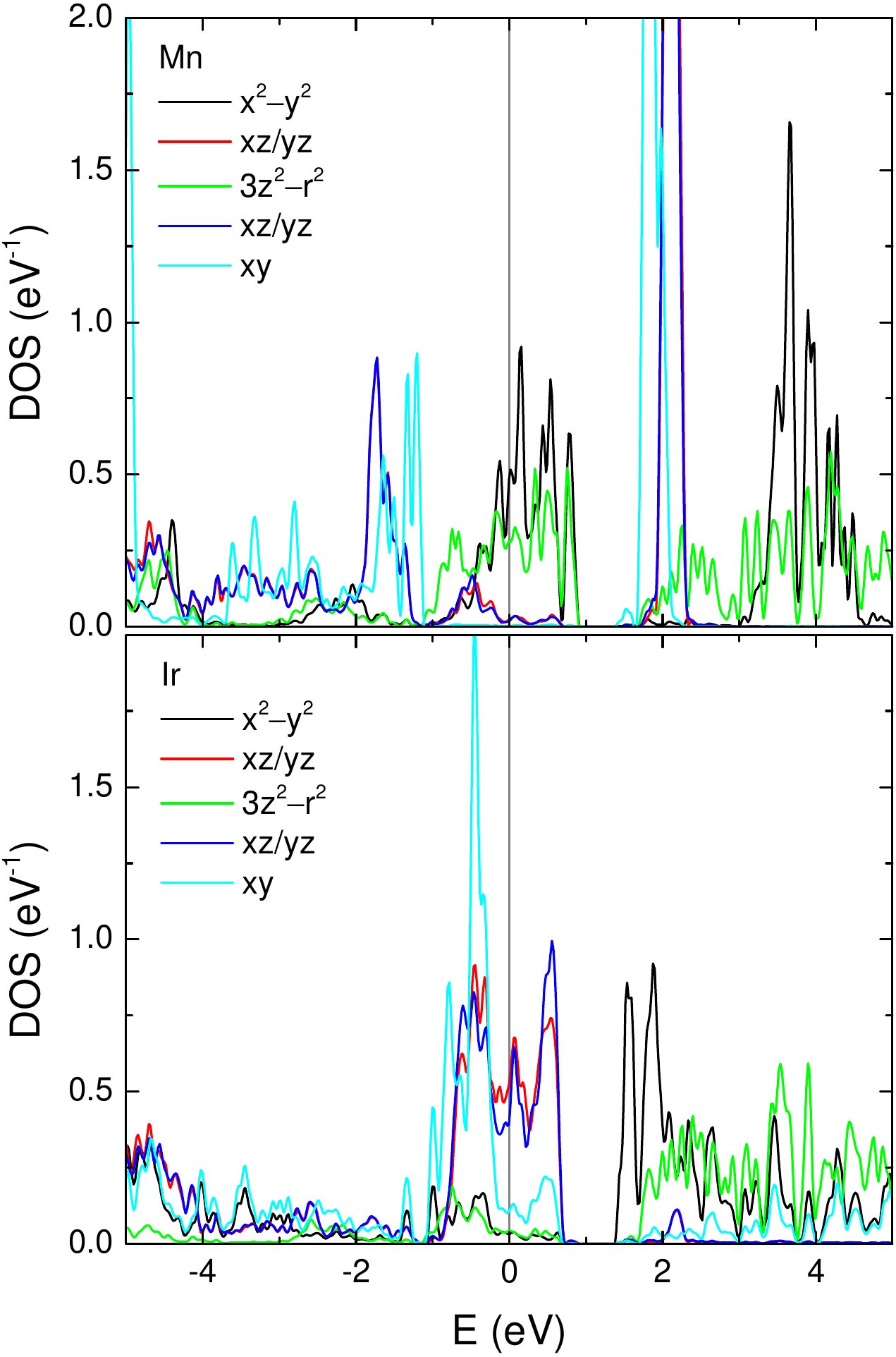}
\caption{Partial density of states of a [SIO]$_1$[SMO]$_1$ superlattice. 
The structural optimization was done without magnetic ordering and the SOC. 
Subsequently, magnetic ordering is introduced with the SOC. 
The $x$ and $y$ axes are rotated from the $a$ and $b$ crystallographic axes by 45 degrees, and $z$ is parallel to the $c$ (out-of-plane) axis.
The Fermi level is set to $E=0$ as indicated by vertical gray lines. }
\label{fig:partialDOS}
\end{center}
\end{figure}

To check this molecular orbital picture, we have plotted the partial DOS of Mn $d$ states and Ir $d$ states in the [SIO]$_1$[SMO]$_1$ superlattice 
in Fig. \ref{fig:partialDOS}. 
The structural optimization was done without magnetic ordering and the SOC. 
Subsequently, magnetic ordering is introduced with the SOC without further structural relaxation. 
We have confirmed that the additional structural relaxation does change the result qualitatively. 
Here, the Fermi level is set to $E=0$. 
One can confirm that Mn $3z^2-r^2$ states are lower than Mn $x^2-y^2$ states and partially filled by electrons. 
Also, for the Ir side, Ir $3z^2-r^2$ states are higher than Ir $x^2-y^2$ states. 
These behaviors are consistent with the molecular orbital argument presented above. 

The molecular orbital picture also explains the difference between [SIO]$_m$[SMO]$_1$ superlattices and [SIO]$_m$[SMO]$_3$ superlattices. 
Because a SMO layer is coupled with two SIO layers in the former, 
the bonding $3z^2-r^2$ orbital here is more stable than that in the latter. 
As a result, for [SIO]$_m$[SMO]$_n$ superlattices, the magnitude of electron transfer from SIO to SMO layers for fixed $m$ decreases with increasing $n$ .

\begin{figure}
\begin{center}
\includegraphics[width=0.85\columnwidth, clip]{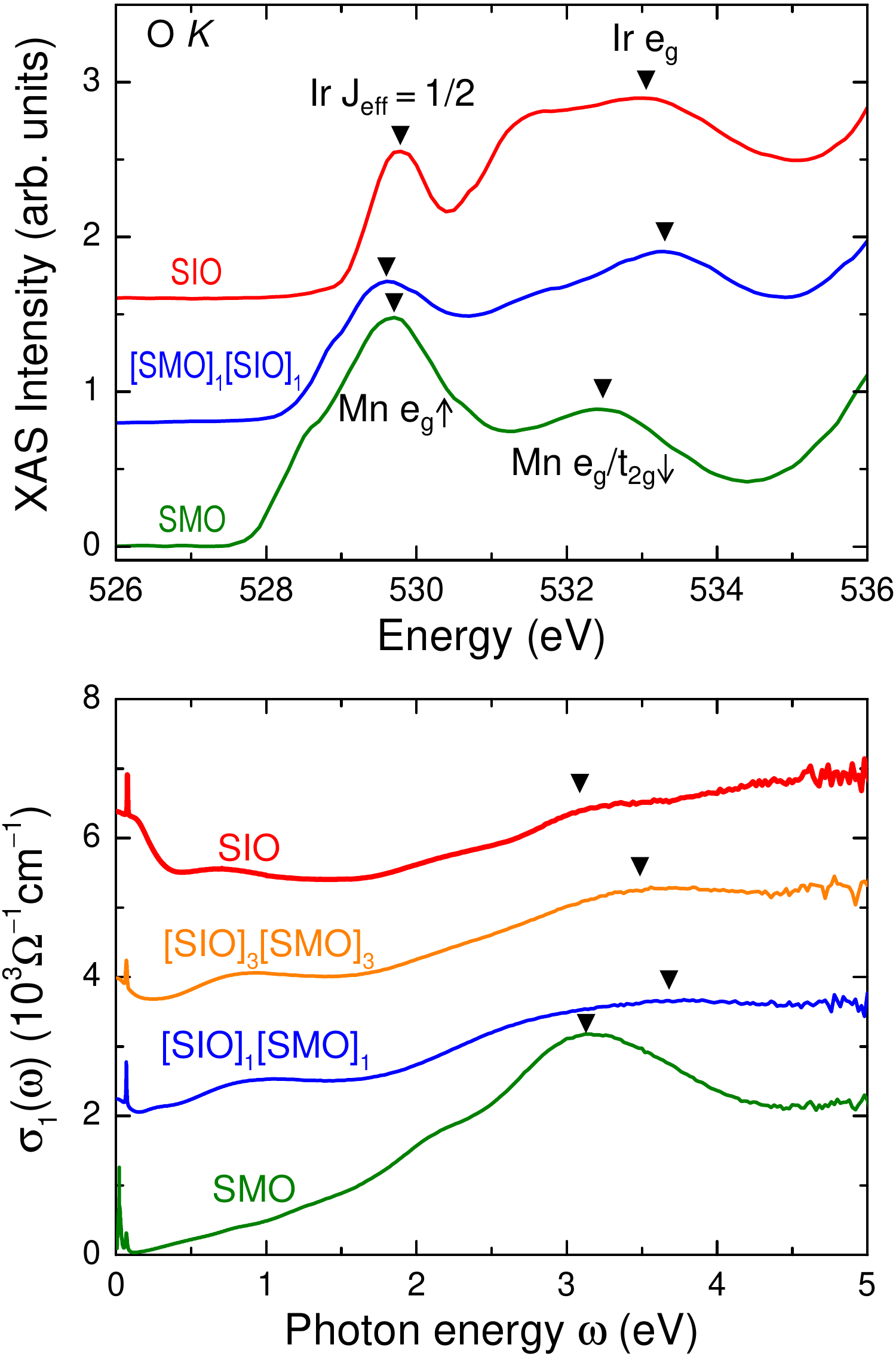}
\caption{XAS (top) and optical conductivity (bottom) spectra of various [SIO]$_m$[SMO]$_m$ superlattices as well as SMO and SIO thin films. 
Note that for both panels, there is a vertical offset for clarity.}
\label{fig:XASandOptCon}
\end{center}
\end{figure}

The molecular orbital picture is consistent with the DOS of this system observed by x-ray absorption spectroscopy (XAS). 
Room temperature XAS spectra of [SIO]$_1$[SMO]$_1$, SIO, and SMO near the O $K$-edge are presented in the upper panel of Fig. \ref{fig:XASandOptCon}. 
The O $K$-edge spectra reflect the transition from O $1s$ core level to unoccupied O $2p$ states. 
Since they are strongly hybridized with Ir and Mn $d$ orbitals, the unoccupied DOS of $d$ orbitals are projected in O $K$-edge spectra. 
The peaks below 530 eV (between 531 and 534 eV) correspond to O $2p$ states hybridized with Ir $J_{eff}=1/2$ and Mn $e_g  \uparrow$ orbitals 
(Ir $e_g$ and Mn $e_g/t_{2g} \downarrow$ orbitals) \cite{Saitoh1995,Matsuno2015}. 
Within the molecular orbital picture, we expect the bonding (antibonding) states of Ir and Mn $e_g$ ($3z^2-r^2$) orbitals to shift to lower (higher) energy. 
Considering the XAS spectrum of the superlattice sample, 
it is clear that the peak near 529.7 eV (533.3 eV) is at a lower (higher) energy than that of either parent material, as indicated with solid triangles. 
This result is in great agreement with the molecular orbital picture that Mn $3z^2-r^2$ (Ir $3z^2-r^2$) orbitals shift to lower (higher) energies.  

In addition, such a molecular orbital picture was also detected in optical spectroscopy. 
The lower panel of Fig. \ref{fig:XASandOptCon} exhibits the real part of optical conductivity $\sigma_1 (\omega)$ at room temperature. 
The solid triangles indicate the peak positions near 3 eV of [SIO]$_1$[SMO]$_1$ and [SIO]$_3$[SMO]$_3$ superlattices as well as SIO and SMO films. 
Following the previous studies on Sr$_2$IrO$_4$,~\cite{Sohn2014} we can assign the peak near 3 eV in SIO as a transition from $J_{eff}=3/2$ to $e_g$ states. 
Note that the corresponding peaks in the superlattices are located at higher energy than those in both parent compounds. 
The molecular orbital picture can explain this peak shift as Ir eg orbitals move to higher energies. 
Thus, both XAS and optical spectroscopic data support the theoretical results, confirming the charge transfer across the SIO/SMO interface.

\begin{figure}
\begin{center}
\includegraphics[width=0.8\columnwidth, clip]{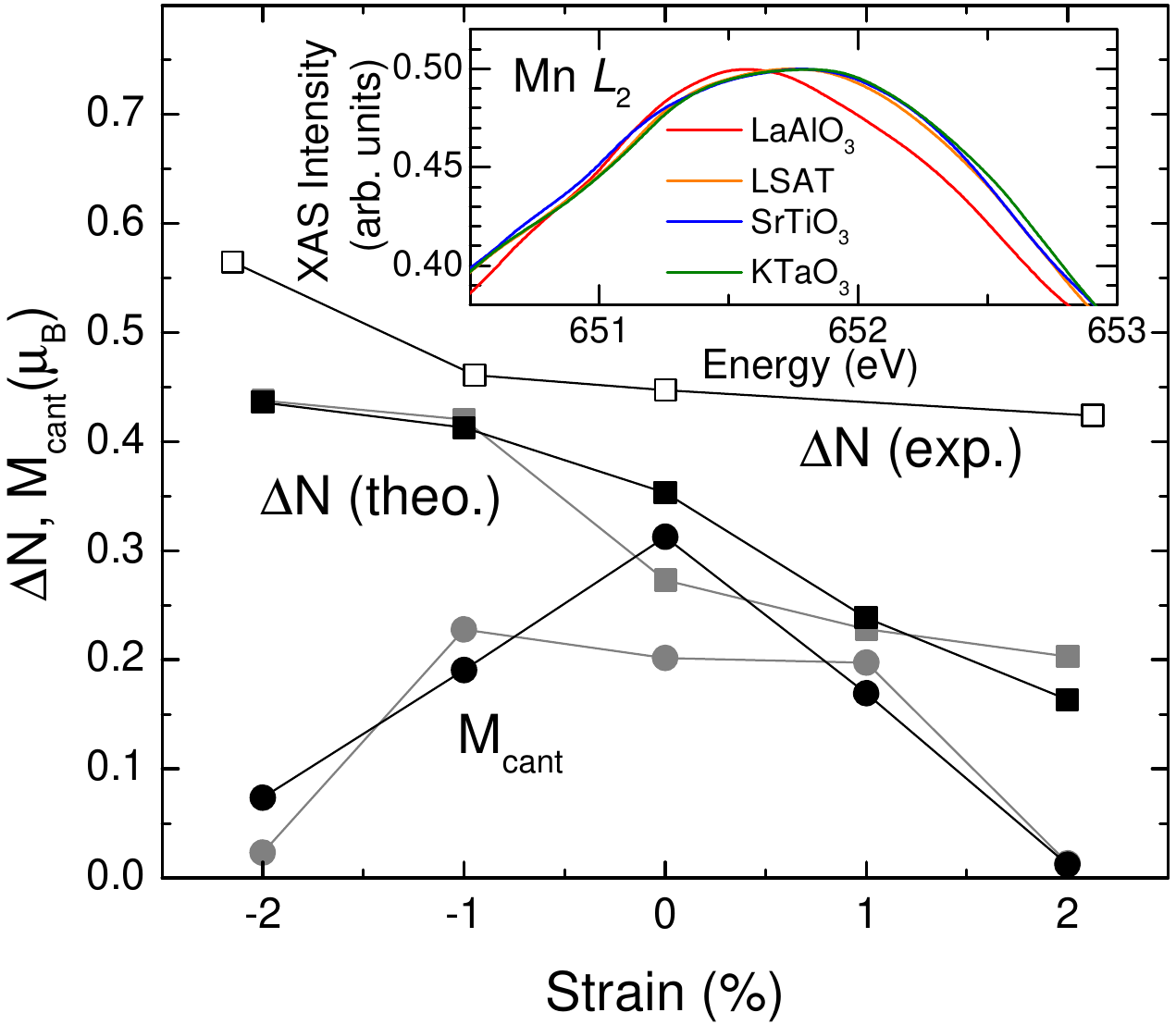}
\caption{Excess charge on Mn sites ($\Delta N$) and the total canted moment ($M_{cant}$) in [SIO]$_1$/[SMO]$_1$ superlattices under various strain states. 
Filled squares and circles are theoretical $\Delta N$, deduced from the change in Mn moments, and the theoretical $M_{cant}$, respectively. 
Black (light) symbols are results of the structural relaxation without the spin polarization and the SOC 
(and the additional structural relaxation with the spin polarization and the SOC with fixed lattice constants obtained previously). 
Open squares are experimental $\Delta N$, obtained from XAS measurements. 
Strain is measured from the lattice constant of SrTiO$_3$. 
Negative (positive) strain is compressive (tensile).
The inset shows the XAS intensity at Mn $L_2$ edge for [SIO]$_1$/[SMO]$_1$ superlattices grown on various substrates: 
LaAlO$_3$ ($-2$~\% strain), LSAT ($-1$~\%), SrTiO$_3$ ($0$~\%), and KTaO$_3$ ($+2$~\%). 
Data was taken at 15 K.}
\label{fig:strain}
\end{center}
\end{figure}

The molecular orbital picture discussed above naturally predicts the following unusual strain effect:  
under the in-plane compressive (tensile) strain, $3z^2-r^2$ molecular orbitals become lower (higher) in energy, 
and as a result, the electron transfer from SIO to SMO is enhanced (suppressed). 
We have repeated the DFT analyses with various strain values. 
The results for the ordered Mn moment are presented in Fig. \ref{fig:strain}. 
We confirmed the expected behavior  
that the Mn moment increases with compressive strain, 
leading to an increased excess electron density $\Delta N$ on Mn. 
Conversely, tensile strain reduces the Mn moment, leading to a reduced $\Delta N$.

We experimentally investigated the strain dependence by synthesizing a series of coherently strained ([SIO]$_1$/[SMO]$_1$)$_z$ 
superlattices on various substrates 
[LaAlO$_3$ ($a = 3.821$~\AA), (LaAlO$_3$)$_{0.3}$(SrAl$_{0.5}$Ta$_{0.5}$O$_3$)$_{0.7}$ (LSAT) ($a = 3.868$~\AA), 
SrTiO$_3$ ($a = 3.905$~\AA), and KTaO$_3$ ($a = 3.988$~\AA)]. 
Note that $z = 64$ for all samples except that on KTaO$_3$, where $z = 22$ to circumvent strain relaxation that is present in thicker superlattices. 
We determined the average Mn oxidation state from the chemical shift of the Mn $L_2$-edge in XAS and observed 
that the Mn oxidation state systematically increases with increasing the in-plane lattice constant through epitaxial strain control. 
This trend is quantified in Fig. \ref{fig:strain} where the strain dependence of $\Delta N$ from stoichiometric SMO is shown.  

The theoretical results on $\Delta N$ and the experimental results agree reasonably well from $\sim -2$~\% to $\sim 0$~\% strain states. 
However, the deviation becomes substantial for large tensile strain $> 0$~\%. 
This might be due to the formation of oxygen vacancies induced by large tensile strain.\cite{Aschauer2013,Petrie2016a,Petrie2016b} 
Note that our samples grown on KTaO$_3$ have a strong tendency to undergo strain relaxation due to the large lattice mismatch 
between KTaO$_3$ and the two constituent materials. 
%
In contrast to the strain dependence of $\Delta N$, 
both superconducting quantum interference device (SQUID) measurements and 
x-ray magnetic circular dichroism (XMCD) measurements indicate that the net moment is suppressed by applying 
both compressive and tensile strain. 
This might be related to the reduction of the spin canting angle under strain. 
As shown in Fig. \ref{fig:strain}, $M_{cant}$ is reduced by strain. 
This reduction by tensile strain is due to the reduction of the rotation angle of the IrO$_6$ octahedron, 
while the reduction by the compressive strain is due to the reduction of  the Ir $d$ occupation.

To summarize, using density functional theory calculations, 
we have investigated the charge transfer in superlattices consisting of SrIrO$_3$ and SrMnO$_3$. 
Despite the nearly identical work function between them and non-polar interfaces, these superlattices show large charge transfer from SrIrO$_3$ to SrMnO$_3$. 
Based on the molecular orbital picture, such large transfer is ascribed to the formation of strong $3z^2-r^2$ bonding orbitals between neighboring Mn and Ir. 
This picture is confirmed by the partial density of states projected on Mn $d$ states or Ir $d$ states. 
The molecular orbital argument also predicts that the amount of electron transfer is controlled by in-plane strain; 
compressive (tensile) strain enhances (suppresses) the amount of electron transfer from Ir to Mn. 
This prediction is readily confirmed by our density functional theory calculations 
along with our XAS and spectroscopic ellipsometry measurements. 
Our work demonstrates a potential new route to control the electronic properties of non-polar oxide heterostructures.

\section*{Acknowledgments}

This work was supported by 
the U.S. Department of Energy,  Office of Science, Basic Energy Sciences, Materials Sciences and Engineering Division.
S. Y. K and T. W. N were supported by the Research Center Program of IBS (Institute for Basic Science) in Korea (IBS-R009-D1). 
Parts of the numerical calculation were performed at the Kavli Institute for Theoretical Physics (KITP), the University of California, Santa Barbara, 
where one of the authors (S. O.) attended the program 
``New Phases and Emergent Phenomena in Correlated Materials with Strong Spin-Orbit Coupling.''
S. O. thanks the KITP, which is supported in part by the National Science Foundation under Grant No. NSF PHY11-25915, for hospitality. 
Use of the Advanced Photon Source, an Office of Science User Facility operated for the US DOE, Office of Science by Argonne National Laboratory, was supported by the US DOE. 
J. N. and H. N. L. thank J. W. Freeland for experimental assistance.






\end{document}